\documentclass[3p,times]{elsarticle}
\usepackage{amssymb,amsmath,graphicx}
\usepackage{color}
\usepackage{lipsum}
\usepackage{mathtools}
\usepackage{cuted}
\usepackage[]{amsmath}
\usepackage{graphics}
\usepackage{amssymb}
\usepackage[]{amsmath}
\usepackage{amsthm}
\usepackage{setspace}
\usepackage{epsfig}
\usepackage{subfigure}
\usepackage{extarrows}

\def\[{\begin{equation}}
\def\]{\end{equation}}

\biboptions{numbers,sort&compress}
\usepackage[dvipdfm,colorlinks,linkcolor=red,anchorcolor=blue,citecolor=green]{hyperref}

\usepackage{amsmath}
\usepackage{times}
\usepackage{anysize}
\marginsize{2cm}{2cm}{0cm}{1.6cm}
\selectfont
\journal{ }
\begin{document}
\begin{frontmatter}
\title{Weakly nonlinear topological gap solitons in Su-Schrieffer-Heeger photonic lattices}
\author{Min Guo$^{1}$, Shiqi Xia$^{1}$, Nan Wang$^{1}$, Daohong Song$^{1}$, Zhigang Chen$^{1,2}$ and Jianke Yang$^{3}$}
\address{$^{1}$
MOE Key Laboratory of Weak-Light Nonlinear Photonics, TEDA Applied Physics Institute and School of Physics, Nankai University, Tianjin, China}
\address{$^{2}$ Department of Physics and Astronomy, San Francisco State
University, San Francisco, CA 94132, USA}
\address{$^{3}$ Department of Mathematics and Statistics, University of
Vermont, Burlington, VT 05405, USA}



\begin{abstract}
We study both theoretically and experimentally the effect of nonlinearity on topologically protected linear interface modes in a photonic Su-Schrieffer-Heeger (SSH) lattice. It is shown that under either focusing or defocusing nonlinearity, this linear topological mode of the SSH lattice turns into a family of topological gap solitons. These solitons are stable. However, they exhibit only a low amplitude and power and are thus weakly nonlinear, even when the bandgap of the SSH lattice is wide. As a consequence, if the initial beam has modest or high power, it will either delocalize, or evolve into a soliton not belonging to the family of topological gap solitons. These theoretical predictions are observed in our experiments with optically induced SSH-type photorefractive lattices.
\end{abstract}
\end{frontmatter}

Topological photonics is currently one of the most active research frontiers \cite{Rev1,Rev2,Rev3} due to its potential to realize robust optical circuitry against disorder, among many proposed applications. In the linear regime, when two materials with different topological invariants (characterized by Zak phase, Berry phase or Chern number) are joined, the bulk-edge correspondence guarantees the existence of topologically protected interface states which exist in the gaps between Bloch bands \cite{Berry,Zak,bulk_edge1,bulk_edge2,bulk_edge3}. Such topological interface states have been introduced and successfully demonstrated in the realm of photonics, exhibiting unconventional electromagnetic wave transport that can overcome disorder and backscattering \cite{Haldane_Raghu2008,Wang2009,Fang2012,Rechtsman2013,Khanikaev2013,Hafezi2013}. However, in the presence of nonlinearity, this bulk-edge connection may break down, and how these topological interface states behave under the nonlinear effects becomes an important question. This question has been investigated in a few topological photonic systems \cite{Zhigang2009,Segev2013,Ablowitz2014,Chong2016,Segev2016,Alu2016,Peschel2019,Rechtsman2020}.
For example, in \cite{Zhigang2009} it was shown that nonlinearity can lead to deformation of topological edge modes in one-dimensional (1D) Su-Schrieffer-Heeger lattices. In \cite{Segev2013}, self-localized wave packets forming topological edge states were reported in the bulk of a 2D nonlinear photonic topological insulator.
In \cite{Ablowitz2014,Chong2016}, families of nonlinear unidirectional edge solitons were theoretically obtained in square or honeycomb lattices of helical waveguides. In \cite{Segev2016}, it was shown theoretically that nonlinear extended edge modes in honeycomb lattices of helical waveguides were always unstable due to modulation instability. In \cite{Alu2016}, it was shown that in a finite bimodal lattice, nonlinearity could induce a topological phase transition and the formation of nonlinear edge states. In \cite{Peschel2019}, it was shown both theoretically and experimentally that topological edge states persist in the nonlinear regime in nonlinear fiber loops, but become linearly unstable above a certain power threshold. Most recently, in \cite{Rechtsman2020}, it was experimentally observed that Floquet topological solitons could form in a photonic lattice modulated periodically along the propagation direction.

The Su-Schrieffer-Heeger (SSH) lattice is one of the simplest topological photonic systems admitting linear topological interface modes \cite{SSH,ZhigangShockley}. In this article, we theoretically and experimentally study nonlinear effects on topological interface states in the SSH lattice established by direct continuous-wave (CW) laser-writing in a photorefractive crystal. We show that, under either focusing or defocusing nonlinearity, there exists a family of topological gap solitons which are linearly stable, and their mode profiles closely resemble the linear topological interface  states. However, such topological solitons only have low amplitude and power and are thus weakly nonlinear, even when the bandgap of the SSH lattice is wide. As a consequence, if the initial beam has modest or high power, it will not evolve into these topological gap solitons. Instead, it will either break up, or evolve into a soliton that does not exhibit the topological feature of the linear mode \cite{Zhigang2020}. These results are predicted theoretically and confirmed experimentally.

Paraxial beam propagation in a photorefractive crystal with a pre-engineered refractive index profile is governed by the equation \cite{Photo_review}
\[
iU_z+\frac{1}{2k}U_{xx}+\frac{k}{n_0}\Delta n(x) U-\frac{kn_0^2r}{2} \frac{E_0}{1+|U|^2}U=0,
\]
where $U(x,z)$ is the envelope of the electric field, $k=2\pi n_0/\lambda$ is the wavenumber, $\lambda$ is the wavelength, $n_0$
is the bulk refractive index, $\Delta n(x)$ is the optically pre-induced refractive index profile, $r$ is the electro-optic coefficient of the crystal, and $E_0$ is the applied DC field. Here, the intensity of the beam has been normalized with respect to the dark irradiance of the crystal. If we measure the transverse direction $x$ in units of $D$, with $D$ being the characteristic scale of the pre-induced photonic lattice, the $z$ direction in units of $2kD^2$, the applied DC field $E_0$ in units of $1/(k^2D^2n_0^2r)$, and define the normalized refractive index profile as $V(x)=2k^2D^2\Delta n(x)/n_0-E_0$, then the above governing equation is non-dimensionalized as
\begin{equation} \label{e:U}
iU_z+U_{xx}+V(x)U+\frac{E_0|U|^2}{1+|U|^2}U=0.
\end{equation}

To study nonlinear effects on topological states, we choose $V(x)$ to be an SSH lattice, whose profile is displayed in Fig.~1a (the blue curve). The characteristic of this SSH lattice is that it is formed by joining two bimodal periodic lattices, connected by a topological defect \cite{Zhigang2020,SSH1,SSH2}. In dimensionless units, each bimodal lattice has period 4.5, with two potential humps of the same height 2 inside each period. Since these two bimodal lattices differ by a proper spatial shift, they have different Berry-Zak phases \cite{Berry,Zak}. In this case, there would be a linear localized topological state at the interface between the two lattices, and the propagation constant of this interface  state lies in the gap between Bloch bands of the lattices \cite{bulk_edge2}. Numerically, we have obtained these Bloch bands as well as the topologically protected linear interface mode, and the results are shown in Fig.~1. It is seen that this topological mode (red curve in Fig.~1a) is symmetric, having peak intensity at the interface, and zero intensity at alternating lattice sites starting from the next neighbors. In addition, it has opposite phase between two neighboring intensity maxima.
\begin{figure}[htbp]
\begin{center}
\includegraphics[width=0.7\textwidth]{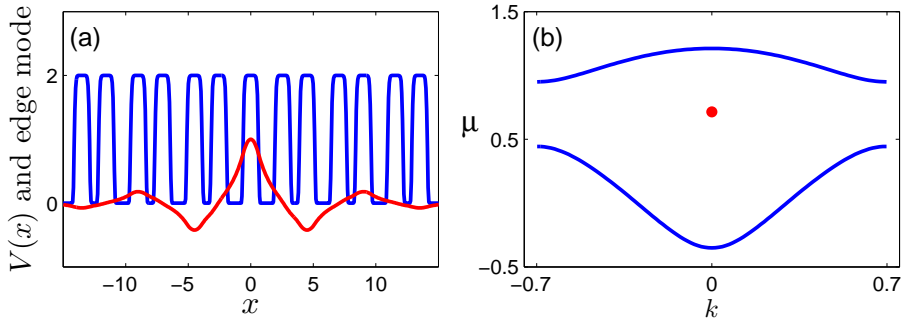}
\caption{(a) An SSH lattice potential $V(x)$ (blue) and its linear interface mode (red).  (b) Bloch bands of the lattices over the Brillouin zone (the red dot is the interface mode in the bandgap).  }
\end{center}
\end{figure}

Now we explore how this topological interface mode is affected by nonlinearity. For this purpose, we look for solitons bifurcating from this linear interface mode. These solitons will be called topological solitons and are of the form
\[
U(x,z)=u(x)e^{i\mu z},
\]
where $u(x)$ is a localized real function satisfying
\[  \label{e:u}
u_{xx}+V(x)u+\frac{E_0u^2}{1+u^2}u=\mu u,
\]
and $\mu$ is the propagation constant. When we choose focusing nonlinearity and set $E_0=7$, this soliton family is shown in the upper rows of Fig.~2. Its power curve is displayed in panel (a), where the power is defined as $P(\mu)=\int_{-\infty}^\infty |u|^2dx$, and its amplitude curve is displayed in panel (d). This soliton family bifurcates to the right side of the linear interface mode's propagation constant toward the first Bloch band. Notice that these amplitudes are all quite low, below 0.3. In addition, their powers are quite low as well (except very close to the band edge). At two points `b, c' of the power curve, the soliton profiles are plotted in panels (b, c), respectively. At very low power (point `b'), the soliton profile closely resembles the linear interface mode of Fig.~1a, which is reproduced as the superimposed red dashed line in Fig.~2b. But at a little higher power (point `c'), the soliton becomes strongly delocalized, indicating that even weak nonlinearity can cause the soliton's delocalization. We have also studied the linear stability of this soliton family by computing their linear-stability spectra, and found that they are all linearly stable. To demonstrate, we plot in panels (e, f) the linear-stability spectra for the two solitons displayed in panels (b, c) respectively. These spectra only contain purely-imaginary eigenvalues, revealing the linear stability of these solitons.

\begin{figure}[htbp]
\begin{center}
\includegraphics[width=0.7\textwidth]{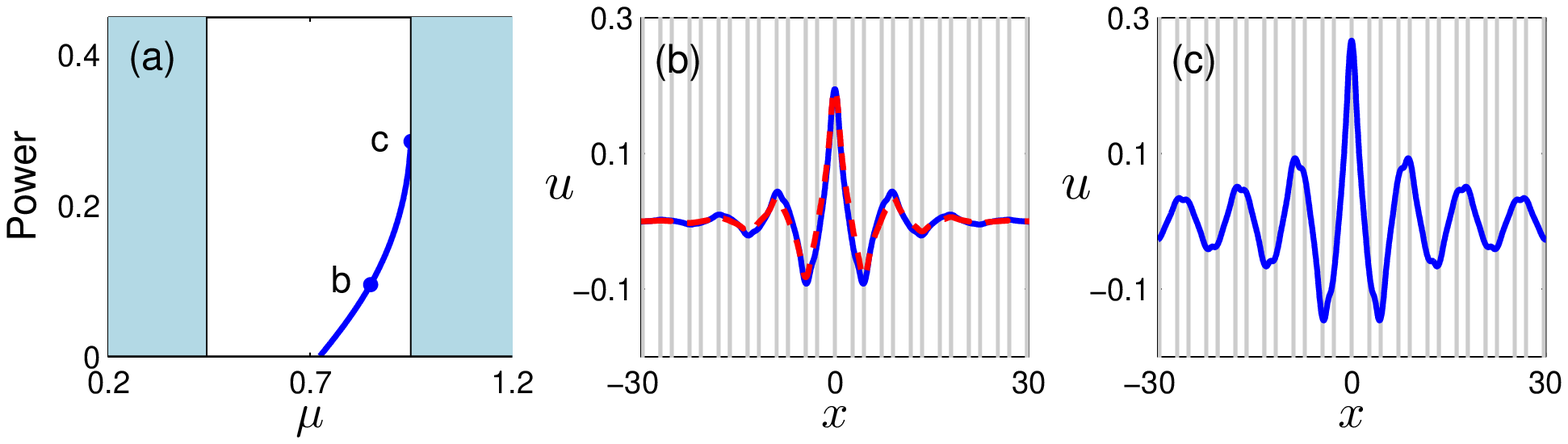}

\includegraphics[width=0.7\textwidth]{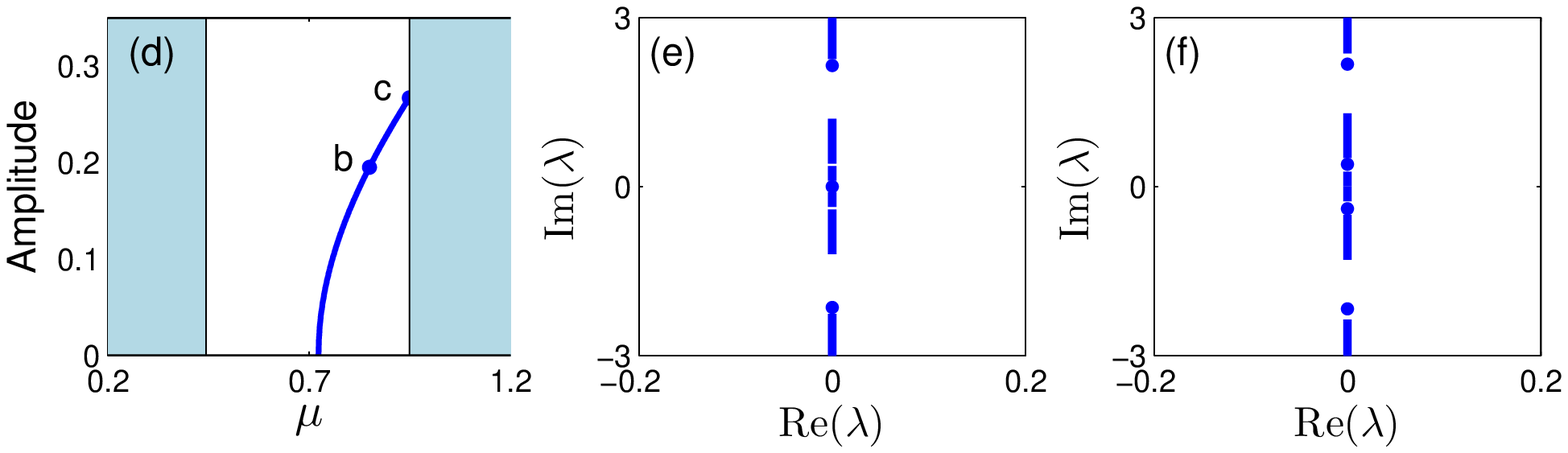}

\includegraphics[width=0.7\textwidth]{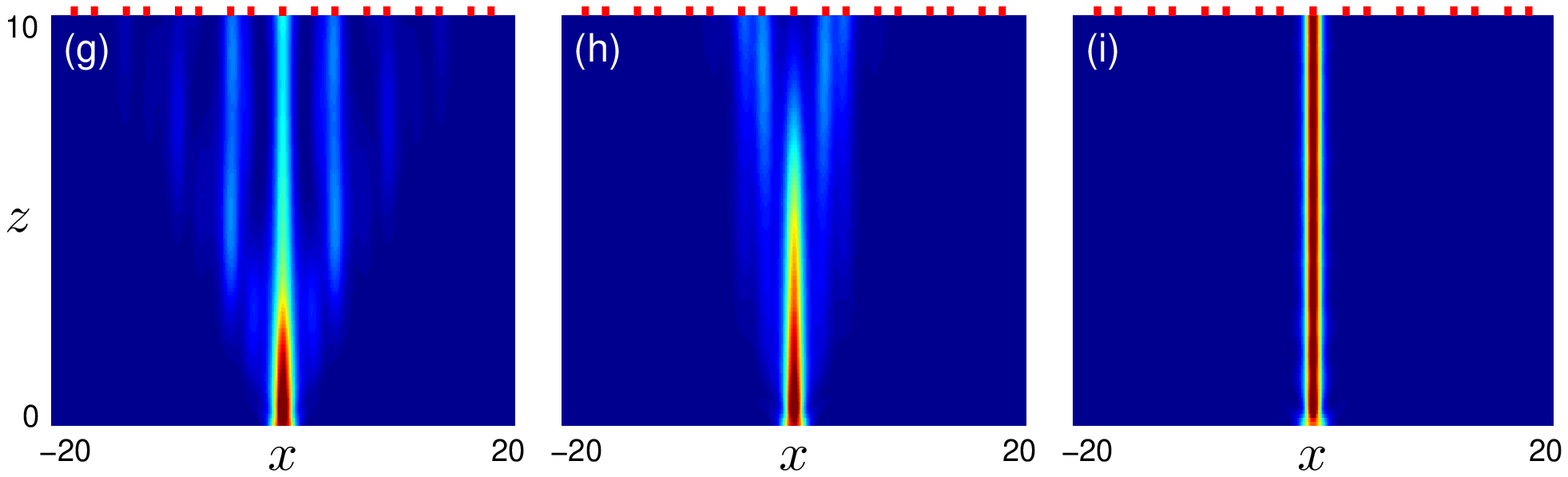}
\caption{Upper two rows: A soliton family bifurcated from the linear topological interface mode of the SSH lattice under focusing nonlinearity with $E_0=7$. (a) Power curve of solitons (shaded regions are Bloch bands). (b,c) Profiles of solitons at points `b,c' of the power curve; vertical stripes represent regions of higher refractive indices of the SSH lattice, and the superimposed red dashed line in (b) is the linear interface mode.  (d) Amplitude curve of solitons. (e,f) Linear-stability spectra of the two solitons in (b,c) respectively. Bottom row: nonlinear evolutions of the Gaussian beam (\ref{U0}) in the SSH lattice for initial amplitude values of $r_0=0.1, 0.4$ and 0.8 respectively; red markers on top of each panel are locations of higher refractive indices of the lattice.}
\end{center}
\end{figure}

The most significant feature of these topological solitons is that their amplitudes and powers are quite low. This means that these solitons are only weakly nonlinear and are not compatible with strong nonlinearity. Thus, we call them ``weakly nonlinear topological gap solitons". If the initial beam carries modest or high power, it will not be able to evolve into these topological solitons. Rather, it will have to either break up or evolve into a nonlinear localized mode not of topological origin. It is noted that under the current focusing nonlinearity, there is indeed a family of solitons residing in the semi-infinite bandgap of the SSH lattice. Those solitons can reach modest and high amplitudes and powers, but they have totally different intensity and phase structures from these weakly nonlinear topological gap solitons.

To test the above evolution predictions, we study the nonlinear evolution of an initial Gaussian beam
\[ \label{U0}
U(x,0)=r_0 \hspace{0.05cm} e^{-x^2/2}
\]
in the SSH lattice, where $r_0$ is the initial amplitude of the beam. At low, medium and high $r_0$ values of 0.1, 0.4 and 0.8, evolutions of this Gaussian beam are obtained by simulating Eq. (\ref{e:U}) and presented in the bottom row of Fig.~2. It is seen that at the low amplitude of $r_0=0.1$, the Gaussian beam evolves into a low-amplitude topological soliton which is very similar to the linear topological interface  state [see panel (g)]. However, at the modest amplitude of $r_0=0.4$, the Gaussian beam breaks up [see panel (h)]. At the higher amplitude of $r_0=0.8$, the beam does self-localize into a stationary soliton state [see panel (i)]. But this soliton is a non-topological soliton residing in the semi-infinite bandgap of the lattice.

\begin{figure}[htbp]
\begin{center}
\includegraphics[width=0.7\textwidth]{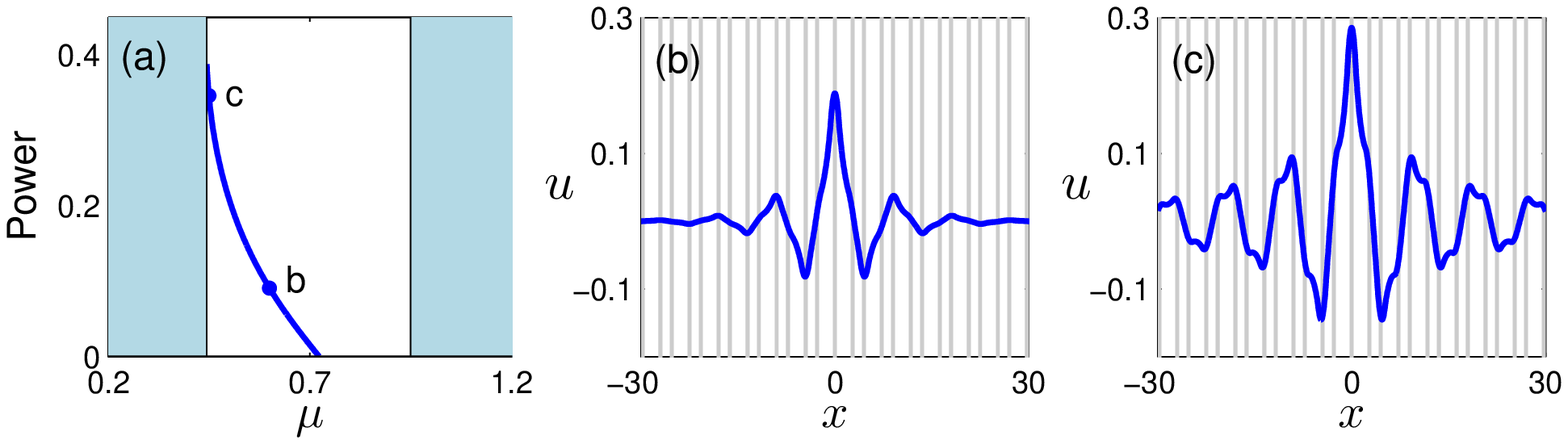}

\includegraphics[width=0.7\textwidth]{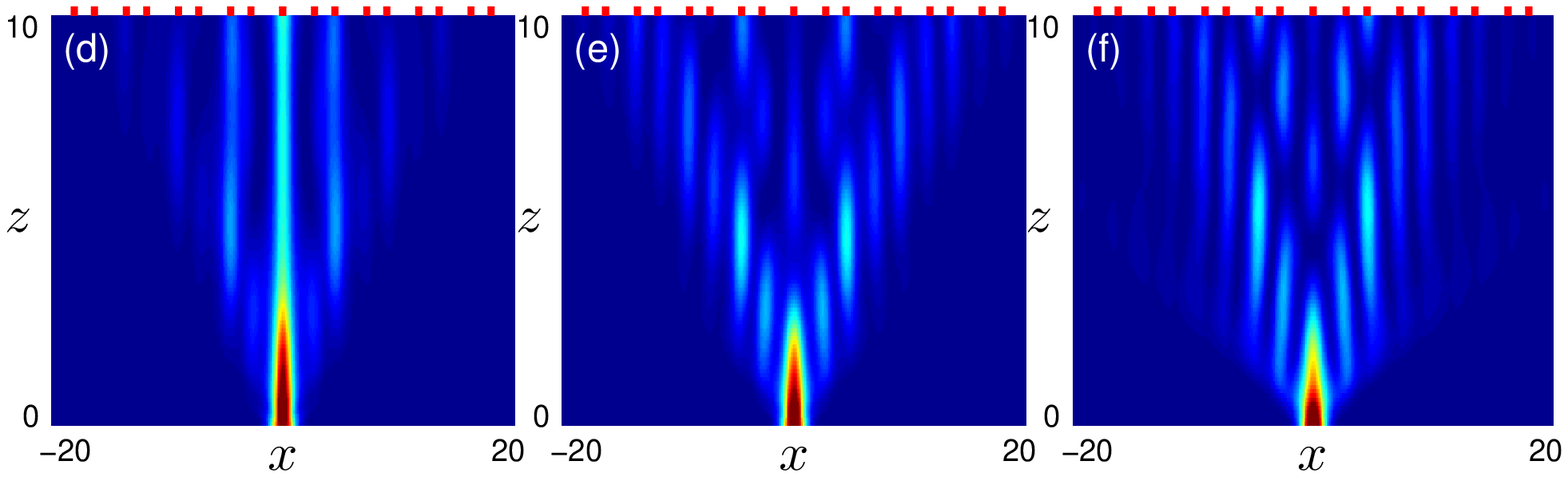}
\caption{Upper row: A soliton family bifurcated from the linear topological interface mode of the SSH lattice under defocusing nonlinearity with $E_0=-7$. (a) Power curve of solitons (shaded regions are Bloch bands). (b,c) Profiles of solitons at points `b,c' of the power curve; vertical stripes represent regions of higher refractive indices of the SSH lattice. Lower row: nonlinear evolutions of the Gaussian beam (\ref{U0}) in the SSH lattice under defocusing nonlinearity for initial amplitude values of $r_0=0.1, 0.4$ and 0.8 respectively; red markers on top of each panel are locations of higher refractive indices of the lattice.}
\end{center}
\end{figure}

What will happen if the nonlinearity is defocusing? For this purpose, we take defocusing nonlinearity by choosing $E_0=-7$. In this case, we can also find a family of topological solitons bifurcating from the linear topological interface mode in the bandgap. The power curve of this soliton family is shown in Fig.~3a, and two representative soliton profiles on the power curve are plotted in Fig.~3(b,c). This power curve bifurcates to the left side of the linear interface mode toward the second Bloch band. Similar to the focusing nonlinearity, these topological solitons also have low amplitude and power and are linearly stable. In addition, as the power increases, the soliton becomes strongly delocalized. Thus, these topological solitons under defocusing nonlinearity are only weakly nonlinear as well, and an initial beam with modest or high power should break up. This expectation is confirmed in the evolution simulations of the initial Gaussian beam (\ref{U0}), which are shown in Fig.~3(d,e,f) for initial amplitude values of $r_0=0.1, 0.4$ and 0.8 respectively.

One may notice that the bandgap in the SSH lattice of Fig.~1a is quite narrow (its width is only about 0.5). Thus, one may wonder if the amplitude and power of its topological solitons would be higher if this bandgap is wider, so that the power curve has more room to grow. To address this question, we increase the height of the normalized refractive index $V(x)$ from 2 to 4. In addition, we increase the spacing contrast between index humps of the lattice. The new lattice profile is plotted in Fig.~4a. For this new lattice, its bandgap has width 1.59, which is three times wider than before. Indeed, its linear topological interface mode is much more localized [see Fig.~4a]. For this wider-gap SSH lattice, we have computed its topological solitons under the same focusing and defocusing nonlinearities with $E_0=\pm 7$ as before, and their power and amplitude curves are shown in Fig.~4(b,c) respectively. It is seen that even though the bandgap is now wider, the power and amplitude of the topological solitons are still low. Thus, they are still not compatible with strong nonlinearities.

\begin{figure}[htbp]
\begin{center}
\includegraphics[width=0.7\textwidth]{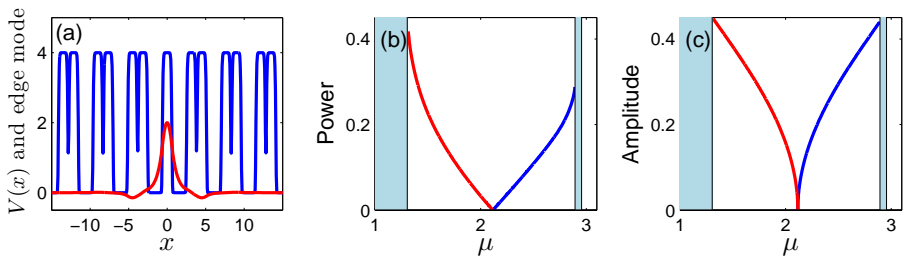}
\caption{(a) An SSH lattice potential $V(x)$ (blue) with wider bandgap and its linear interface mode (red).  (b, c) Power and amplitude curves of solitons bifurcated from the linear topological interface mode of the SSH lattice in (a) under focusing (blue, with $E_0=7$) and defocusing (red, with $E_0=-7$) nonlinearities. Shaded regions in (b,c) are Bloch bands.}
\end{center}
\end{figure}

Our experimental results of nonlinear effects on the topological interface states are summarized in Fig.~5. The SSH lattice with an interface similar to that shown in Fig.~1a is established by direct CW-laser-writing method in a nonlinear SBN crystal \cite{Zhigang2020}, since such an SSH lattice with an interface topological defect cannot be established by conventional multi-beam interference based optical induction method {\cite{SSH,ZhigangShockley}. The lattice written is shown in the bottom panel of Fig.~5a, with a period about 45$\mu$m. The bias electric field used in lattice writing is 2.4 kV/cm, and the resulting lattice index variation is about $4.36\times 10^{-4}$. The input stripe beam (Fig.~5b) is launched straightly into the interface waveguide (marked by a red dot in the top panel of Fig.~5a). This input beam, after 2cm of linear propagation, diffracts normally without the lattice (Fig.~5c), but evolves into a linear topological interface state through the lattice (Fig.~5d). Notice that this observed interface state is similar to that shown in Fig.~1a. After the nonlinearity is turned on by applying an electric field across the crystal ($880$V/cm for the self-focusing nonlinearity or $-800$V/cm for the self-defocusing nonlinearity), when the input power of the probe beam is weak (2.13$\mu$W for the self-focusing and 0.37$\mu$W for the self-defocusing cases), the output intensity pattern still resembles that of the linear topological state (see Fig.~5e and Fig.~5g). This indicates that, at low power (weak nonlinearity), the probe beam has evolved into a topological gap soliton, as analyzed in our theory. On the other hand, under strong nonlinearity (by increasing the power of the probe beam
to 5.9$\mu$W for self-focusing and 6.81$\mu$W for self-defocusing cases), the output intensity pattern changes dramatically. Specifically, under self-focusing nonlinearity, the output exhibits strong localization into the initially excited interface waveguide (Fig.~5f), which corresponds to generation of a nonlinear Tamm-like surface state (or discrete semi-infinite-gap soliton) not of topological origin \cite{Zhigang2009,Zhigang2020}. In contrast, under strong self-defocusing nonlinearity, the output pattern becomes strongly delocalized and spreads into the bulk (Fig.~5h). These experimental results under weak and strong nonlinearity conditions agree well with the theoretical predictions shown in Fig.~2 and Fig.~3.

\begin{figure}[htbp]
\begin{center}
\includegraphics[width=0.7\textwidth]{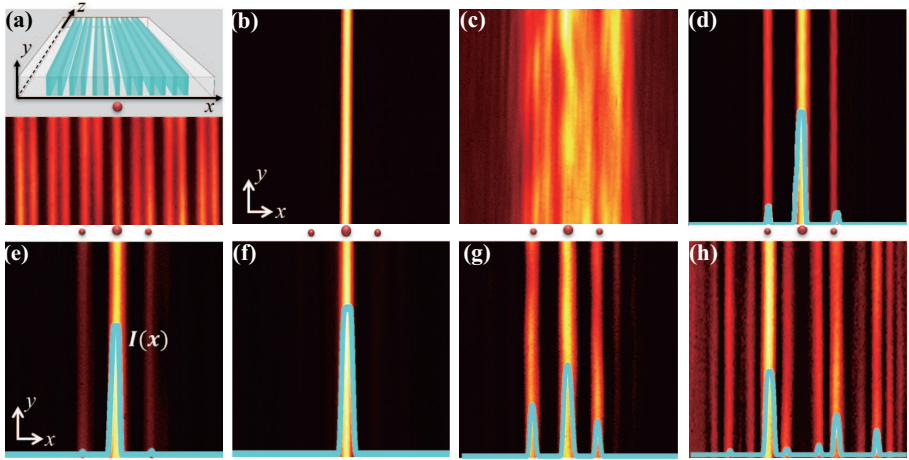}
\caption{Experimental results of linear and nonlinear interface single-channel excitation in a photonic SSH lattice. (a) Schematic (top) and experimentally established (bottom) SSH photonic lattice by cw-laser-writing, where the red dots mark the positions of the interface and its next-nearest lattice sites. (b) Probe beam at input. (c, d) Linear output of the probe beam without (c) and with (d) the lattice. (e, f) Nonlinear output under weak (e) and strong (f) self-focusing nonlinearity. (g, h) Nonlinear output under weak (g) and strong (h) self-defocusing nonlinearity. All intensity patterns are taken at input/output $(x, y)$ transverse planes, as illustrated in (b, e). The blue curves superimposed in (d-h) are the plots of corresponding intensity profiles along the $x$-direction. }
\end{center}
\end{figure}

In our theoretical model (\ref{e:U}), the nonlinearity is saturable, which is appropriate for photorefractive crystals used in our experiment \cite{Photo_review}. In silicon waveguides with a femto-second-laser-written SSH lattice, the nonlinearity is Kerr (cubic) rather than saturable. However, the topological solitons in our saturable model (\ref{e:U}) are all weakly nonlinear, in which case the saturable nonlinearity becomes effectively cubic. Thus, our results in this article should be valid for femto-second-laser-written waveguides as well. This could well be a general feature of nonlinear topological states bifurcated from their linear counterparts, regardless the type of nonlinearity \cite{Zhigang2020}.

In summary, we have theoretically and experimentally studied nonlinear effects on topologically protected interface states in the photonic Su-Schrieffer-Heeger (SSH) lattices. We have shown that, under either self-focusing or self-defocusing nonlinearity, topological gap solitons would bifurcate out from the linear topological mode of the SSH lattice, provided that the nonlinearity is weak. As a result, initial beams with modest or high powers would either break up or evolve into nonlinear modes without topological features. This inability of the SSH lattice to support strongly nonlinear topological gap solitons suggests that the SSH lattice and strong nonlinearity may be incompatible, and thus the robustness and topological protection of interface  states under nonlinear excitation merit further investigation. Our results may also have ramifications for other nonlinear topological systems beyond optics.

\vspace{0.1cm}
Funding. National Science Foundation (DMS-1910282), Air Force Office of Scientific Research (FA9550-18-1-0098), National Key R\&D Program of China (2017YFA0303800), National Natural Science Foundation of China (11922408, 91750204, 11674180).

\end{document}